\pdfoutput=1

\documentclass{article}
\usepackage[affil-it]{authblk}
\usepackage[utf8]{inputenc}
\usepackage{graphicx}
\usepackage{siunitx}
\usepackage[margin=1in]{geometry}
\usepackage[sorting=none]{biblatex}
\title{A TPC-based tracking system for a future Belle II upgrade}

\author[1]{Andreas Löschcke Centeno}
\author[1]{Christian Wessel}
\author[1]{Peter M. Lewis \thanks{email: lewis@physik.uni-bonn.de}}
\author[2]{Oskar Hartbrich}
\author[1]{Jochen Kaminski}
\author[3]{Carlos Mariñas}
\author[2]{Sven Vahsen}

\affil[1]{\normalsize University of Bonn, Institute of Physics, Nu\ss{}allee 12, 53115 Bonn, Germany}
\affil[2]{\normalsize University of Hawaii, Department of Physics and Astronomy, 2505 Correa Rd., Honolulu, HI 96822, USA}
\affil[3]{\normalsize University of Valencia - CSIC, Instituto de Fisica Corpuscular (IFIC), Spain}

\newcommand\snowmass{\begin{center}\rule[-0.2in]{\hsize}{0.01in}\\\rule{\hsize}{0.01in}\\
\vskip 0.1in Submitted to the  Proceedings of the US Community Study\\ 
on the Future of Particle Physics (Snowmass 2021)\\ 
\rule{\hsize}{0.01in}\\\rule[+0.2in]{\hsize}{0.01in} \end{center}}

\begin{document}

\maketitle

\snowmass

\begin{abstract}
In the next decade, intensity frontier experiments will require tracking systems that are robust against high event and background rates while maintaining excellent tracking performance. We develop a first conceptual design of a tracking system for a hypothetical future experiment---here imagined as a successor to Belle II---built around a time projection chamber (TPC) with high resolution readout. This choice necessitates a significant expansion of the silicon vertex detector as well as a new fast timing layer. We simulate the performance of such a system in the Belle II simulation framework, probe its major technical challenges, and demonstrate that such a system is suitable for projected luminosities at the next generation of intensity-frontier colliders. 
\end{abstract}

\section{introduction}
Intensity frontier experiments, particularly those at the $B$ Factories, require high-precision tracking for fairly low-momentum tracks in the presence of high event and beam-induced background rates. Such experiments typically rely on drift chambers to measure helical segments of charged tracks with a minimal material budget. Operational experience in the early stages of Belle II, the current state-of-the-art $B$ Factory experiment, has shown that the drift chamber technology may be reaching its limit due to high occupancy. To address this issue, we have developed and simulated a first conceptual design for a Time Projection Chamber (TPC)-based tracking system for a hypothetical future $B$ factory experiment, which we outline here. Further technical details of the TPC can be found in Ref.~\cite{andreas}.

For convenience and concreteness, we suppose that SuperKEKB and Belle II have been superseded by ultra-high luminosity upgrades, operating with the same beam energies but with five times the maximum instantaneous luminosity ($5\times 6.5\times 10^{35} \si{cm^{-2}.s^{-1}}$), and assume that the proposed tracking system is surrounded by existing Belle II components. However, in principle the concept is equally suitable for a Belle II upgrade or a future, unrelated intensity-frontier experiment. 
\section{A tracking concept}
In the ultra-high luminosity scenario, we assume that the geometry is constrained by the existing PID and electromagnetic calorimetry systems. With this constraint, we have considered three competing proposals. The first is an upgraded drift chamber. We consider this solution unsuitable given the current challenges of operating Belle II's drift chamber (CDC) at SuperKEKB well below design luminosity. The second option is a full silicon tracker. We have conducted preliminary simulations of such a system and found that it significantly degrades the $p_T$ resolution of tracks due to increased multiple scattering. This, coupled with the intrinsic cost and structural difficulties of such a system, suggests that such a system is not suitable. The third option is a TPC-based tracker, which should provide a significant reduction in occupancy because it is a true three-dimensional detector, while drift chambers are effectively two-dimensional. However, a TPC tracker has some intrinsic limitations: it cannot provide a trigger signal as the CDC does, and event pileup can be very high due to the long electron drift time. We will see that these problems are easily overcome, and we base our conceptual design on such a system. 

\begin{figure}
\begin{center}
    \includegraphics[width=10cm]{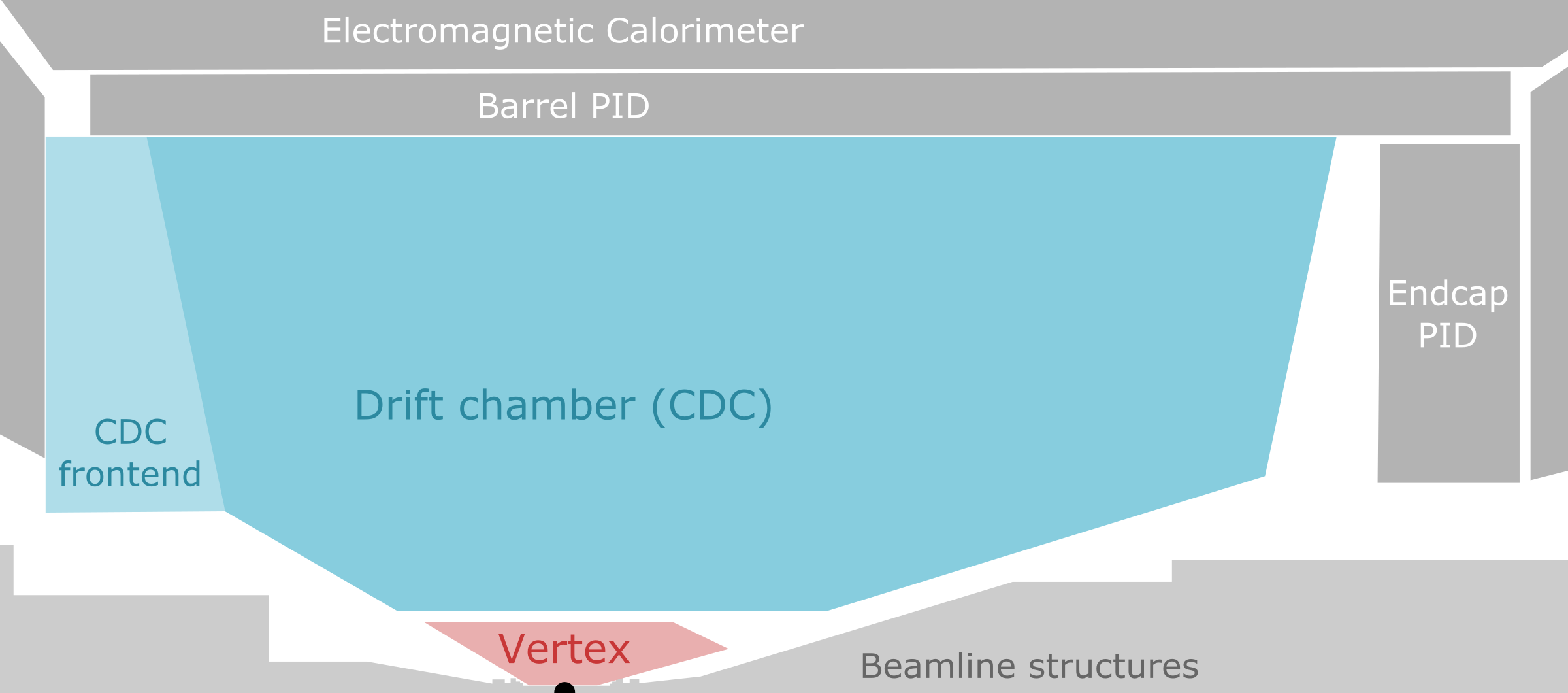}
    
    \vspace{0.5cm}
    
    \includegraphics[width=10cm]{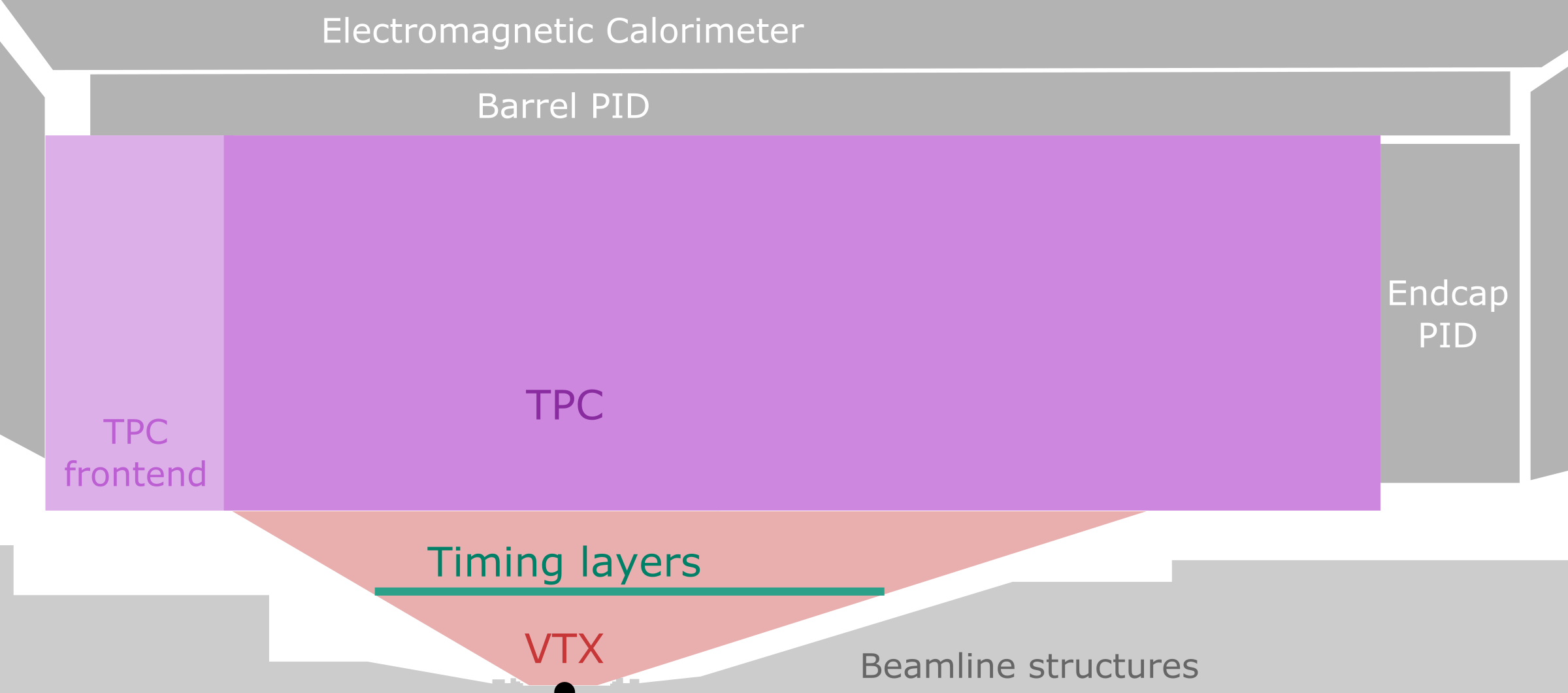}
  \caption{A schematic cross-section of the (top) Belle II and (bottom) hypothetical high-luminosity upgrade tracking systems. The electron beam travels along the bottom of the figures to the right and collides with the left-going positron beam at the black semicircle.}
\label{fig:schematic}
\end{center}
\end{figure}

Our basic conceptual design consists of three components, illustrated in Fig.~\ref{fig:schematic}. First, the current CDC is replaced by a TPC with a single drift volume and readout on the backward endcap. Second, the current silicon vertex detectors are replaced with a new detector, which we base off of the VTX upgrade proposal. In order to maintain an annular cylinder geometry for the TPC, we extend the VTX from a radius of $\SI{13.5}{cm}$ to $\SI{44}{cm}$. Third, we insert a multilayer fast timing detector~\cite{hartbrich}, possibly silicon, at $r=25$~cm or $r=45$~cm in order to replace the triggering role of the CDC and additionally provide particle identification (PID) via time-of-flight (TOF) for low-$p_{T}$ tracks.

\subsection{Technical requirements}
We consider the tracking system concept viable if it achieves similar performance to the CDC in realistic conditions. We perform our performance tests at either of two scenarios: at SuperKEKB full luminosity (the ``$\times 1$" scenario) or at projected ultra-high luminosity (``$\times 5$").

\subsection{Technical challenges} \label{sec:challenges}

We target our early studies at answering the most challenging questions raised by the tracking TPC concept, including:
\begin{enumerate}

    \item \textit{Event pileup}: Can tracks be reliably associated with unique events despite a high degree of event overlap?
    \item \textit{Background pileup}: Can beam-induced background hits be reliably identified and rejected despite integrating over $\SI{60}{\micro\meter}$ of drift time?
    \item \textit{Ion density}: Will the ion backflow rate be small enough that it does not significantly degrade the tracking performance?
    \item \textit{Continuous readout}: Can all of these be accomplished with a continuous readout design (without gating), which is necessitated by the high physics event rates?
    \item \textit{Diffusion}: Can the TPC match the tracking performance of the CDC by using a high number of space-time points to overcome the limitations of diffusion?
\end{enumerate}
We consider demonstrations of positive answers to these questions to constitute a proof-of-principle for the tracking TPC project. In order to focus on these key challenges instead of more basic technical design optimizations, we borrow heavily from work already done for the ILD TPC~\cite{ild_gridpix}. In the next section, we explain our assumptions and design. 

\subsection{Proof-of-principle TPC design}\label{sec:tpc}
Our basic design consists of a single gas volume of length \SI{242}{cm} with high-resolution readout tiling at the backward endcap. We do not use any form of ion gating and we assume that an external detector will provide the trigger, which will be used to isolate discrete event ``windows" in the continuous data. 

We assume that the TPC uses atmospheric pressure T2K gas mixture~\cite{t2k}. We use MAGBOLTZ~\cite{magboltz} to determine the a drift field ($289$~V/cm) that minimizes the drift time of electrons ($v_D=\SI{7.89}{\centi\meter/\micro\second}$, leading to a maximum drift time of roughly $\SI{30}{\micro\second}$ for a maximum drift length of $\SI{242}{\centi\meter}$). From the same simulation we find that the longitudinal and transverse diffusion coefficients are $\sigma_L=\SI{200}{\micro\meter/\cm}^{0.5}$ and $\sigma_T=\SI{84}{\micro\meter/\cm}^{0.5}$.

For charge amplification and readout, we take inspiration from the GridPix system proposed for use in LCTPC. GridPix consists of an array of $55\times \SI{55}{\micro\meter}$ pixels with a MICROMEGAS mesh mounted onto the surface. For our purposes, this technology presents a number of advantages: first, the small pixels and direct mapping between amplification cells and pixels constitute essentially a best-case resolution scenario. Second, in theory such a sensor can be operated in \textit{binary readout} mode in which each individual hit represents exactly one electron and consists only of the pixel ID and a relative threshold-crossing time. This can dramatically reduce the data throughput, which we anticipate to be a significant technical challenge at ultrahigh luminosities and with continuous readout. Finally, it is very easy to implement such a detector in the digitization simulation.

\section{Simulation}\label{sec:simulation}
We use the Belle II event and detector simulation to generate primary ionization for physics events and beam-induced backgrounds. However, due to the event-based nature of the simulation framework, these simulated events must be merged in external code, in such a way that the effects of event overlay, drift, and diffusion are encapsulated. 

\subsection{Event overlay}\label{sec:overlay}
We generate a collection of events of each of the expected event types (including Bhabha scattering and $ee \to \{\Upsilon(4S)$,\ $u\bar{u},\ d\bar{d},\ s\bar{s},\ c\bar{c},\ \gamma\gamma,\ \mu\mu(\gamma), \tau\tau,\ eeee,$ and $ee\mu\mu$\}) in proportion to their known cross sections. We then simulate ionization in the TPC using Geant4 within the basf2 framework~\cite{b2_sim}. 

We then distribute these events randomly in time and add Gaussian smearing independently in the longitudinal direction and in the transverse plane to approximate diffusion depending on the total drift distance of each primary charge. After drift and diffusion, we bin into 3D voxels (nominally $55\times55\times\SI{55}{\micro\meter\cubed}$) to approximate the amplification digitization in a GridPix-like detector operating in binary readout mode. 

Finally, we center the time window on a single $\Upsilon(4S)$ event and discard hits from outside a $\SI{30}{\micro\second}$ window approximating the total drift time of a single event. An example of a single overlay is shown in Fig.~\ref{fig:eventoverlay}.

\section{Studies}\label{sec:studies}
Our proof of concept consists of simulation studies targeting the primary concerns of Sec.~\ref{sec:challenges}, using the simulation techniques described in Sec.~\ref{sec:simulation}.

\subsection{Event pileup}\label{sec:event_pileup}
Using the event overlay procedure described in Sec.~\ref{sec:overlay} with 100 simulated events, we find that the mean number of physics background tracks per triggered event is approximately $9$. This is somewhat smaller than the typical number of $\Upsilon(4S)$ tracks in an event (roughly $11$). The background tracks are almost entirely from Bhabha scattering (Fig.~\ref{fig:eventoverlay}). All physics background tracks originating from collisions apart from the triggered one should be straightforward to identify and remove: not only will they fail to point to the interaction point, but they will have a transverse diffusion width inconsistent with their relative $z$ position. 

Even at extrapolated luminosities, we find that the number of tracks overlapping with the triggered event is small, and that these are relatively easily identified and removed. Consequently, we do not consider event pileup to be a major technical barrier to the TPC tracking concept. 

\subsection{Background pileup}\label{sec:background_pileup}
Beam-induced background hits are largely produced by \textit{microcurlers}. These are low-$p_T$ electrons produced when low-energy photons interact in the TPC volume. These electrons are fully captured in the parallel electric and magnetic fields, and have far longer path lengths in the TPC volume than typical physics tracks. Consequently, a single microcurler typically deposits far more ionization energy than a physics track. This, coupled with the huge abundance of low-energy beam-induced photons, means that beam-induced backgrounds dominate the ionization rate at both the $\times1$ and $\times5$ luminosity scenarios. 

\begin{figure}
	\centering
	\includegraphics[width=\textwidth]{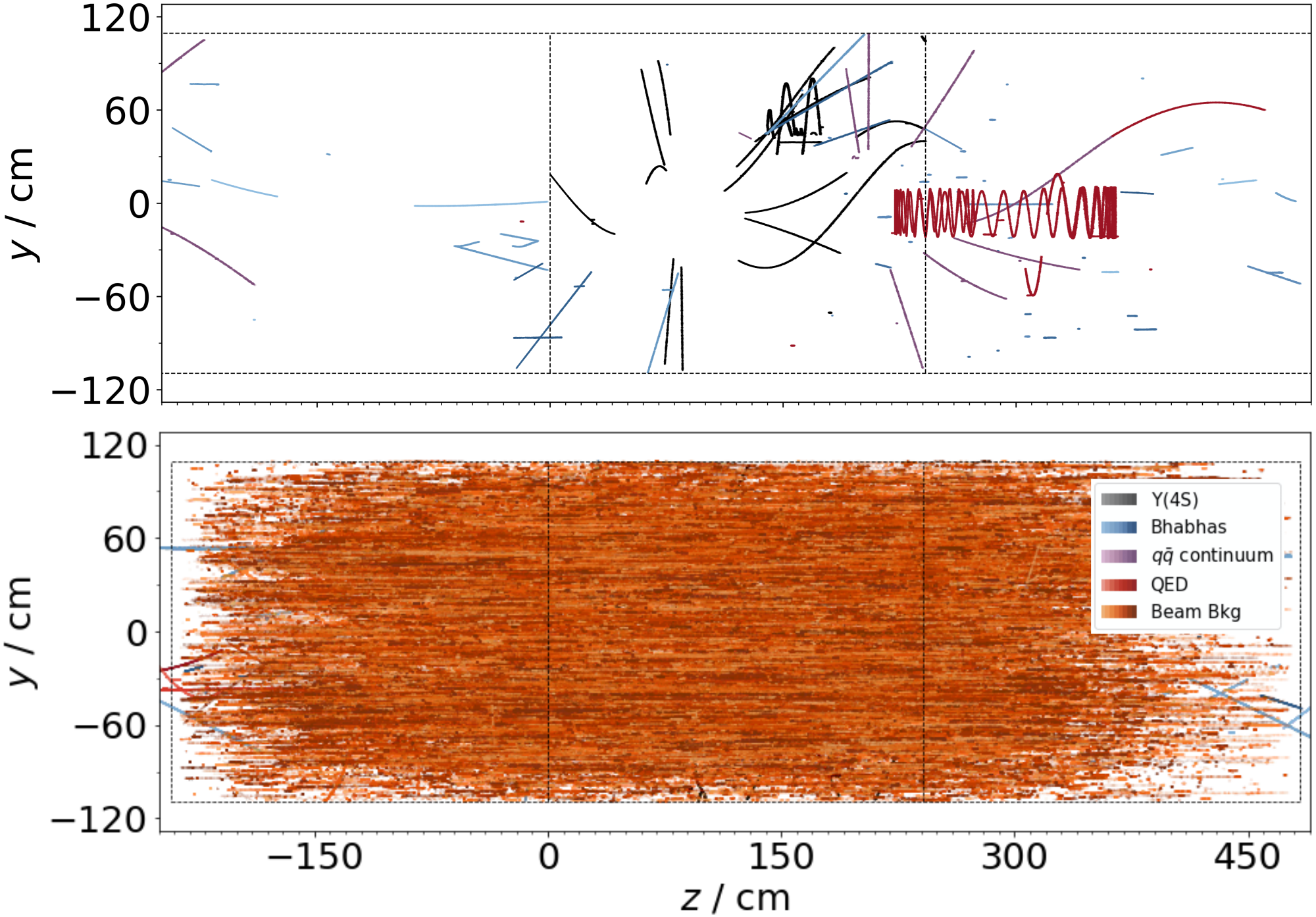}
	\caption{Typical overlays without (top) and with (bottom) $\times5$ beam-induced background hits. Three TPC volumes are shown, with the central volume between $z=0$ and $242$~cm centered on an $\Upsilon(4S)$ decay and corresponding to the read-out event. An external trigger on the $\Upsilon(4S)$ event is assumed. Although modest overlaps from physics backgrounds (primarily Bhabha scattering) exist, the total ionization rate is dominated by microcurlers from beam-induced backgrounds.}
	\label{fig:eventoverlay}
\end{figure}

\begin{figure}
	\centering
	\includegraphics[width=\textwidth]{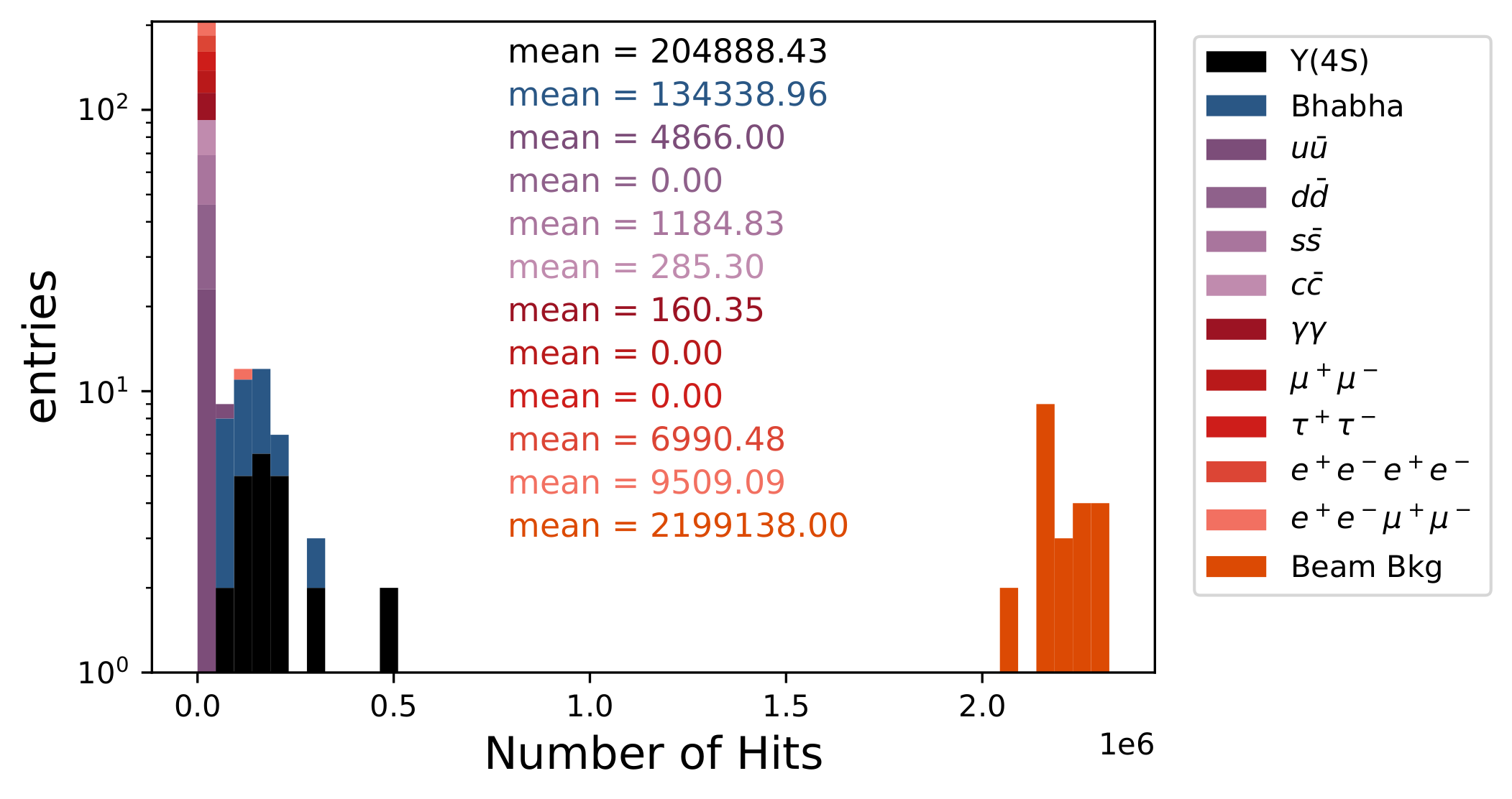}
	\caption{A histogram of the total ionization charge (equivalent to detector hits) deposited in the TPC volume for a number of externally triggered $\Upsilon(4S)$ events. The approximate mean number of hits from $\Upsilon(4S)$ events is $200\times10^3$, from Bhabha events is $130\times10^3$, from all other physics backgrounds is $6\times10^3$ and from beam-induced backgrounds ($\times1$) is $2200\times10^3$. }
	\label{fig:nhits}
\end{figure}

Figure~\ref{fig:eventoverlay} shows typical overlaid events with and without beam-induced backgrounds ($\times5$ scenario). Figure~\ref{fig:nhits} shows the typical deposited charge in triggered $\Upsilon(4S)$ events by source. In both figures it is clear that beam-induced background hits dominate over physics sources. In fact, this estimate might be quite optimistic: SuperKEKB uses continuous injection~\cite{skb}, and associated backgrounds are not included in the beam simulation. Assuming that the ultra-high luminosity upgraded accelerator similarly uses continuous injection, we would expect the TPC to typically integrate over the backgrounds produced by multiple individual injections. Consequently, the total beam-induced background ionization rate may be significantly higher than seen in our simulations.

However, even if we multiply the mean beam-induced background hit rate in the $\times5$ scenario by a safety factor of 5 to account for probable injection backgrounds, we find a typical voxel occupancy of $2\times 10^{-4}$, assuming relatively large $200\times 200\times\SI{200}{\micro\meter\cubed}$ voxels. Furthermore, microcurlers are typically quite distinctive, depositing small rings of high charge density on the readout plane, and consequently a large number of them can be easily identified and rejected, perhaps even before readout. We find that simply rejecting clusters of hits that are isolated in $2\times \SI{2}{\centi\meter\squared}$ cells can reject a majority of microcurlers. This scenario mimics an easily implemented self-trigger mask that requires hits near the edges of a chip to accept an external readout trigger. Further front-end measures to prevent even reading out  microcurlers are possible. 

Given the low total occupancy of beam-induced background hits and the relative ease of identifying and rejecting them early, we conclude that they do not constitute a major technical barrier to the TPC concept. However, this certainly presents one of the more challenging and interesting considerations for the design of the readout system. 

\subsection{Ion density}\label{sec:ion_density}
A second potential challenge posed by the high ionization rates is \textit{ion backflow} (IBF). We initially assume that no gating is used, in order to preserve continuous readout. Therefore, the mean IBF rate is determined by the primary ionization rate, the ion drift velocity, the gain, and the fraction of backflowing ions from the amplification. Using an educated guess of $1\%$ backflow at a gain of 2000, and using measurements of the mobilities of the primary ions in the T2K gas~\cite{t2k}, we find a typical ion charge density comparable to that of other tracking TPCs. However, this number is certainly optimistic: injection backgrounds are not simulated and we expect them to be very high since continuous injection will presumably be used. Furthermore, the TPC ionization rate is dominated by microcurlers, most of which come from beam-induced low-energy photons. Therefore, the ion density depends strongly on our SuperKEKB beam background simulation that may not be suitable for the future upgrade scenario. Therefore, we cannot rule out the possibility that high backflowing ion densities will substantially degrade the tracking performance.

Considerable further work is necessary to simulate and measure the IBF and to determine its effects on the tracking performance. The mean ``time occupancy" of $\Upsilon(4S)$ events in the $5\times$ scenario is roughly $15\%$, which suggests that gating is not feasible, as was assumed.   

\subsection{Tracking performance}\label{sec:tracking_performance}
The key question concerning tracking performance is whether the increase in the number of spatial hit points (up to the theoretical maximum; the number of ionizations) can win out over diffusion when determining tracking parameters. Here we focus primarily on the transverse momentum $p_T$. We use a set of $1000$ simulated muons in discrete bins of $p_T$ and distributed uniformly in $\theta$ over the acceptance of the TPC. 

\begin{figure}
	\centering
	\includegraphics[width=\textwidth]{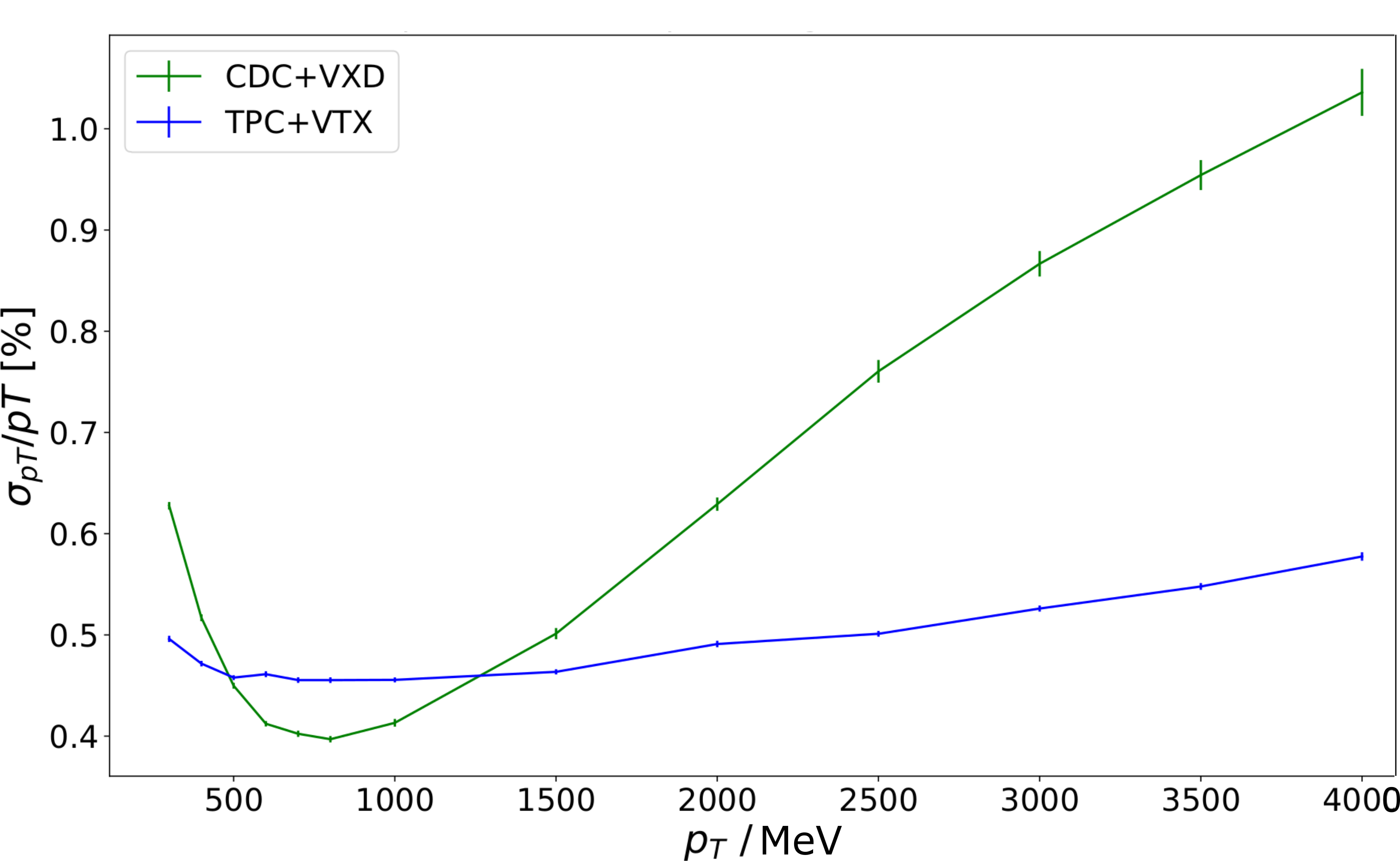}
	\caption{A comparison of the average $p_T$ resolution for the CDC+VXD and TPC+VTX tracking systems.}
	\label{fig:pTRes_PTCol}
\end{figure}

Figure~\ref{fig:pTRes_PTCol} shows the simulated $p_T$ resolution for the current (CDC+VXD) vs. proposed (TPC+VTX) tracking systems averaged over $\theta$. The vertical offset is due to multiple scattering, which is largely determined by the material budget of the vertex detectors, which for the VTX is highly speculative. In principle, due to significant thinning of each layer, the TPC+VTX should be able to achieve comparable or lower levels of multiple scattering compared to the CDC+VXD.

The linear slope in $p_T$, due to position measurement resolution, is far shallower for the TPC compared to the CDC. This is due to the large number of hit points in the TPC, rendering its effective point resolution far superior to the hit resolution in the CDC. We conclude that the TPC is capable of matching or surpassing the tracking resolution of the CDC. However, the tracking performance in the critical range $p_T<\SI{1}{\giga\eV}$ depends more strongly on the amount of material in the VTX than it does on the differences between the CDC and TPC.  

\begin{figure}
	\centering
	\includegraphics[width=\textwidth]{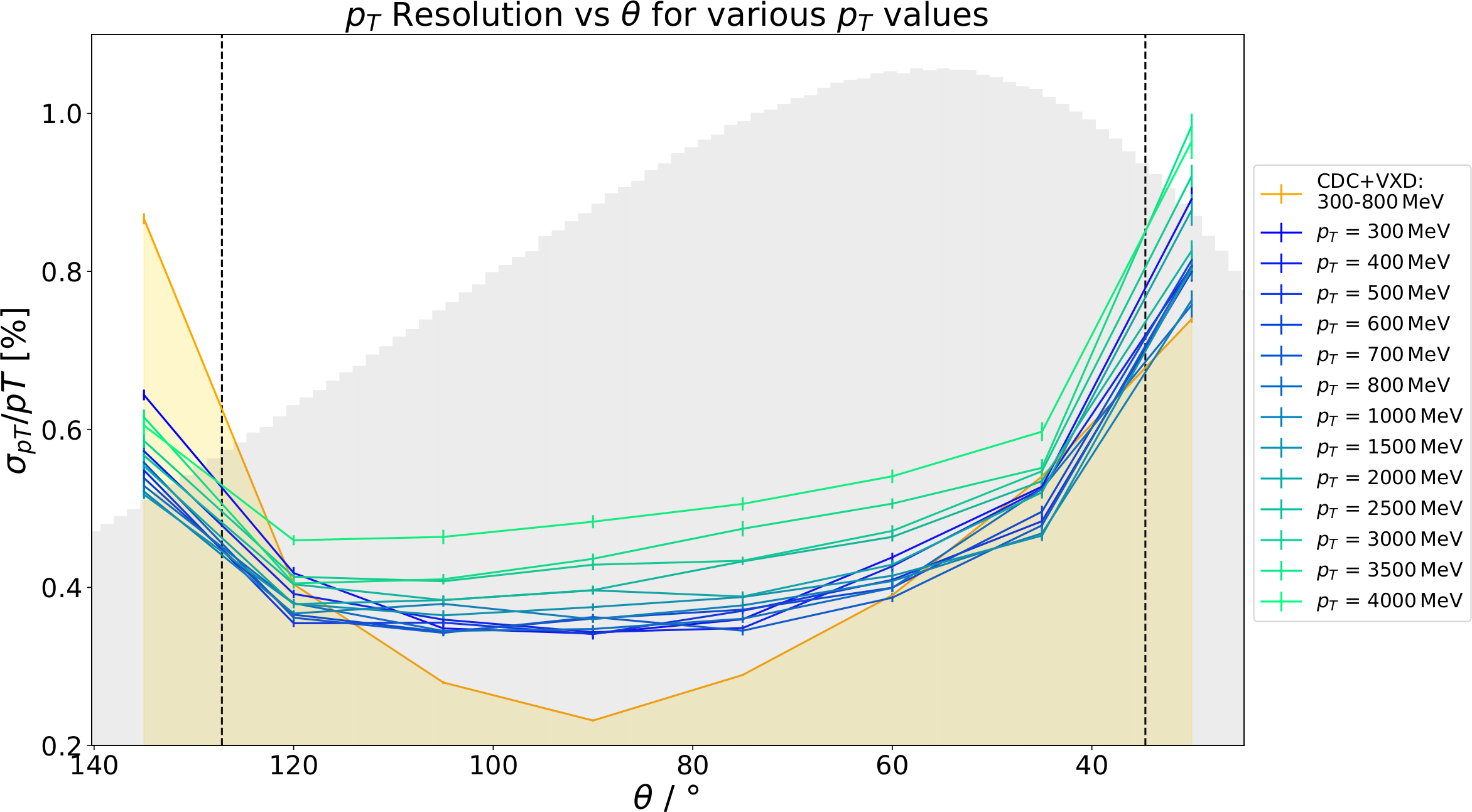}
	\caption{A comparison of the $p_T$ resolution for CDC+VXD compared to TPC+VTX vs. polar angle $\theta$. The gray histogram in the background indicates the distribution of tracks in $\Upsilon(4S)$ decays. The dashed lines marks the boundary between the TPC endcaps and barrel.}
	\label{fig:pTRes_PT}
\end{figure}

With transverse tracks ($\theta=\SI{90}{\degree}$, with a drift length of $85$~cm), the CDC+VXD has better $p_T$ resolution than the TPC+VTX, due to a combination of multiple scattering in the VTX and diffusion in the TPC (Fig.~\ref{fig:pTRes_PT}). Additionally, at this angle the tracks have their minimum path length in the TPC, leading to a minimum in the number of hits. However, the TPC+VTX resolution degrades far less quickly for shallower angles due to an increase in the number of hit points. This benefit compared to CDC tracking is amplified in the backward region where diffusion in the TPC is at its minimum. 

\begin{figure}
	\centering
	\includegraphics[width=\textwidth]{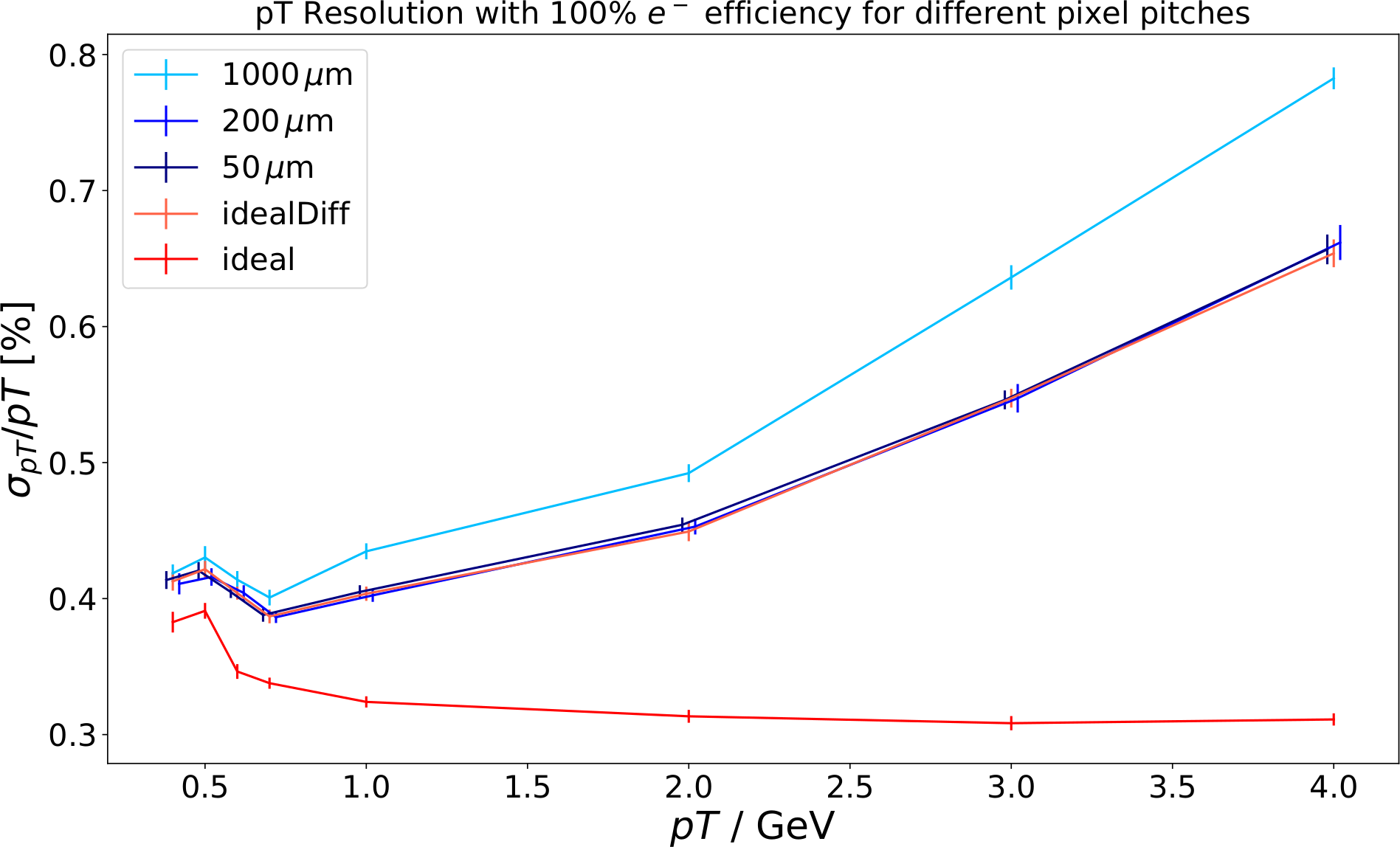}
	\caption{A comparison of the TPC+VTX tracking resolution for various readout scenarios. \textit{Ideal} is for a TPC with no diffusion and perfect resolution, \textit{idealDiff} adds realistic diffusion, and the other three scenarios indicate simulated pixel pitches.}
	\label{fig:ptRes_E100}
\end{figure}

In order to determine the contributions to the effective point resolution, we study how the $p_T$ resolution depends on diffusion and pixel pitch (Fig.~\ref{fig:ptRes_E100}). We find that diffusion dominates the effective resolution, and consequently there is no benefit to decreasing the pitch far below the typical diffusion width (roughly $\SI{1}{\milli\meter}$). Under the assumption that it is desirable to minimize the channel count, we select the largest pixel pitch that shows now significant degradation in $p_T$ compared to the ideal case: $\SI{200}{\micro\meter}$.

\subsection{Triggering}\label{sec:triggering}
\begin{figure}
	\centering
	\includegraphics[width=\textwidth]{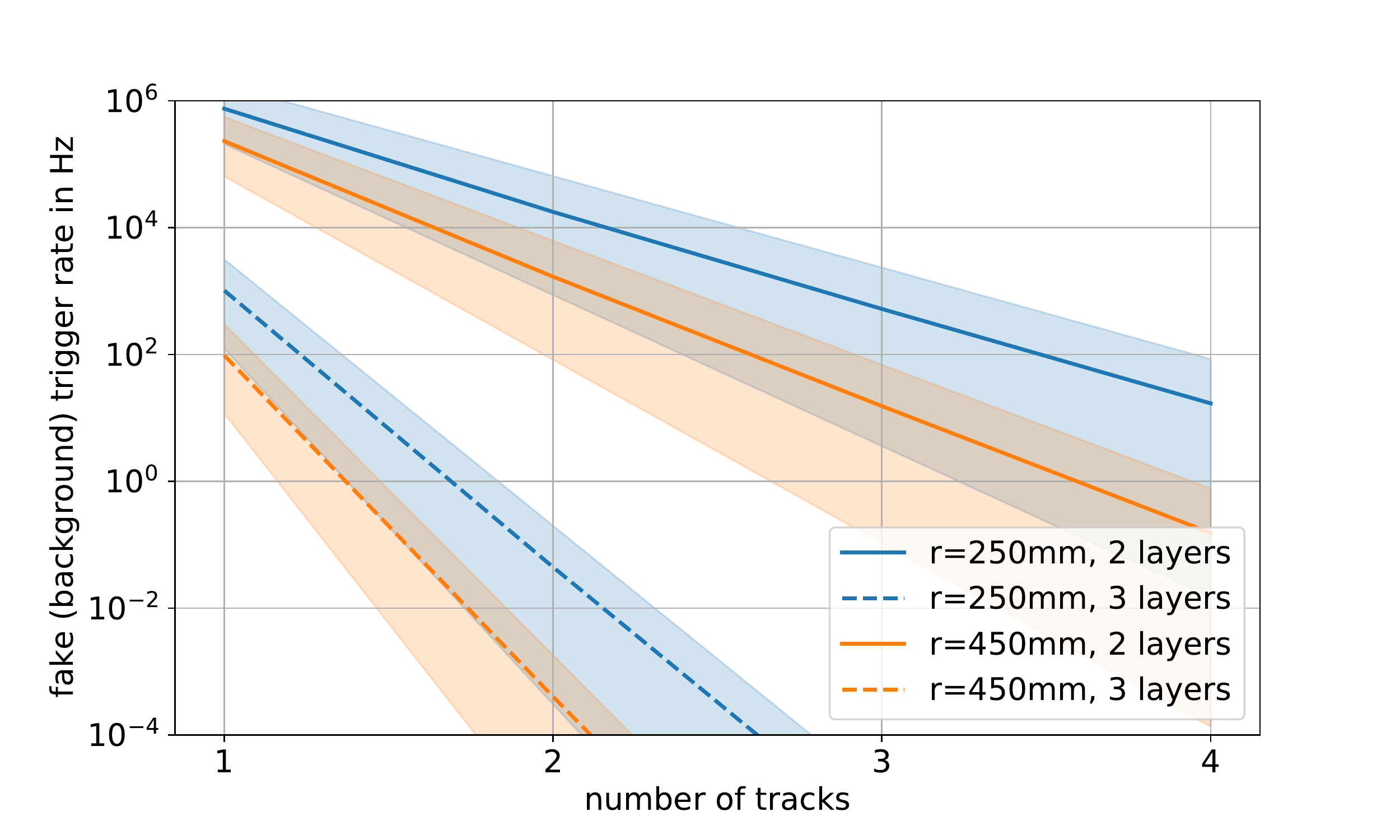}
	\caption{Simulation of the trigger performance (in the $\times1$ scenario) of the fast timing detector in both 2- and 3-layer versions, and at positions $r=25$ and $45$~cm. }
	\label{fig:timing_track_trigger}
\end{figure}

For the timing layer, we envision multiple layers of fast silicon sensors between the beam pipe and the TPC. For concreteness, we assume that this sensor uses the same technology as in the STOPGAP proposal~\cite{hartbrich}, which uses these sensor to increase the hermiticity of the Belle II PID system outside of the tracking volume. As designed, this detector has a time resolution better than $\SI{100}{\pico\second}$, and we place it in the middle of the VTX (at $r=\SI{25}{\centi\meter}$) or between the VTX and the TPC (at $r=\SI{45}{\centi\meter}$). Triggers are generated based on coincidence between layers. Based on simulation with a 3-layer detector at 25cm, a two-track trigger will have a fake rate of less than 1Hz (Fig.~\ref{fig:timing_track_trigger}). 

\begin{figure}
	\centering
	\includegraphics[width=\textwidth]{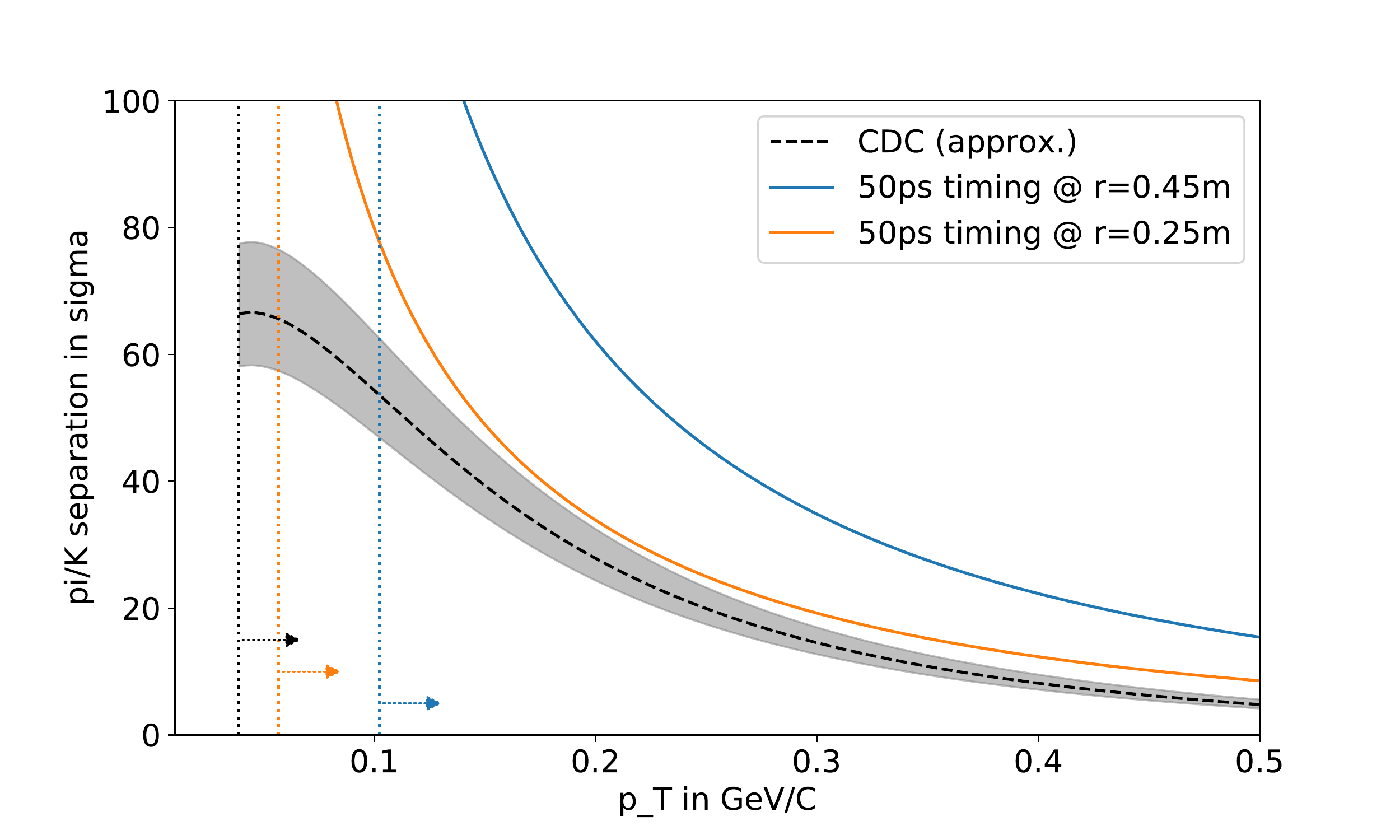}
	\caption{Simulated $K/\pi$ separation power vs. $p_T$ for a mutilayer fast-timing detector located at $r=25$ and $\SI{45}{\centi\meter}$. The dashed vertical lines indicate the $p_T$ acceptance of each detector.}
	\label{fig:tof_pid_lin2}
\end{figure}

A promising capability of such a timing detector is that it provides PID information based on time-of-flight (TOF). This compensates for the loss of $dE/dx$ PID information from the inner CDC volume abandoned to the VTX. Fig.~\ref{fig:tof_pid_lin2} shows the $K/\pi$ discrimination power of the simulated timing detector acting as a TOF PID detector. For the $r=25$~cm scenario, excellent separation down to $\SI{50}{\mega\eV}$ is possible, with improved performance relative to the CDC. 

From this preliminary work, we conclude that the combination of a TPC with a multi-layer timing detector at $r=\SI{25}{\centi\meter}$ is capable of matching or surpassing the triggering and low-$p_T$ PID capabilities of the CDC. 


\subsection{Additional studies}\label{sec:additional}
Further studies, detailed in Ref.~\cite{andreas}, suggest that the $p_T$ resolution is fairly insensitive to electron efficiency, that the maximum data throughput would be comparable to the current VXD ($7$~GB/s), assuming front-end hit filtering), and that the pixel dead-time has a negligible effect on electron efficiency. In aggregate, these studies indicate no additional cause for serious doubt about the prospects of TPC-based tracking at the next generation of intensity-frontier experiments. 
\section{Conclusions}\label{sec:conclusions}
Based on these studies, we conclude that a gas TPC-based tracking system may be viable for intensity frontier experiments like the hypothetical ultra-high luminosity upgrade to Belle II. Our design relies heavily on the capabilities of the GridPix sensor, particularly the association of a single pixel with a single electron with low ion backflow and excellent 3D resolution. The primary difficulty of such a system is the cost of tiling these sensors over the $> \SI{3}{\square\meter}$ endplate. However, we find that binary readout with relatively large pixels ($200\times\SI{200}{\micro\meter}$) is sufficient to meet the performance objectives of the tracking system, which decreases the channel count and data throughput of the system, perhaps offsetting some of the costs. The timing layer proposal is an essential part of the TPC-based tracking system, but is perhaps independently desirable as a source of low-$p_T$ PID discrimination.

The sole major outstanding technical question is whether the backflowing ion densitities will remain tolerable after factoring in real-world beam-induced backgrounds, including those caused by injection. Considerable work is needed to answer this question definitively, including the development of prototypes, testbeam campaigns, and a mature design and simulation of the upgraded accelerator and interaction area. For now, we can only conclude that our preliminary simulations do not rule out the tracking TPC concept on the basis of ion density.

\printbibliography

@mastersthesis{andreas,
    author = {Löschcke Centeno, A.},
    institution = {University of Bonn},
    title = {First Conceptual Design and Studies fo a Tracking Time Projection Chamber for the Belle II Experiment},
    url = "https://docs.belle2.org/record/2631/files/BELLE2-MTHESIS-2021-073.pdf",
    year = 2021
}

@article{hartbrich,
    author = "Hartbrich, Oskar and others",
    eprint = "2203.04847",
    archivePrefix = "arXiv",
    primaryClass = "hep-ex",
    month = "3",
    year = "2022"
}

@article{ild_gridpix,
	author = {Kaminski, J.},
	title = {Large Area Coverage of a TPC Endcap with GridPix Detectors},
	DOI= "10.1051/epjconf/201817402001",
	journal = {EPJ Web Conf.},
	year = 2018,
	volume = 174
}

@article{b2_sim,
author = {Kim, D.-Y. and others},
year = {2017},
title = {The simulation library of the Belle II software system},
volume = {898},
journal = {Journal of Physics: Conference Series},
doi = {10.1088/1742-6596/898/4/042043}
}

@article{skb,
    author = {Ohnishi, Y. and others},
    title = "{Accelerator design at SuperKEKB}",
    journal = {Progress of Theoretical and Experimental Physics},
    volume = {2013},
    number = {3},
    year = {2013},
    doi = {10.1093/ptep/pts083}
}

@article{t2k,
author = {Cortez, A. F. Z. and others},
title = "{Experimental ion mobility measurements for the LCTPC Collaboration}",
journal = {NIM-A},
volume = {936},
pages = {451-452},
year = {2019},
doi = {https://doi.org/10.1016/j.nima.2018.11.049}
}

@misc{magboltz,
title = {Monte Carlo simulation of electron drift and diffusion in counting gases under the influence of electric and magnetic fields},
author = {Biagi, S F},
doi = {10.1016/S0168-9002(98)01233-9},
journal = {NIM-A},
volume = {421},
year = {1999}
}

\end{document}